\documentclass[manuscript, screen]{acmart}
\renewcommand\footnotetextcopyrightpermission[1]{} 

\AtBeginDocument{%
  }
\usepackage{appendix}

\setcopyright{none}
\acmConference[CHI '25]{Conference on Human Factors in Computing Systems}{April 26--May 1, 2025}{Yokohama, Japan}

\begin{document}

\title{Resisting AI Solutionism through Workplace Collective Action}

\author{Kevin Zheng}
\email{kyzheng@umich.edu}
\orcid{0009-0001-7870-3916}

\author{Linda Huber}
\email{ludens@umich.edu}
\orcid{0000-0003-3726-8433}

\author{Aaron Stark}
\email{astark@umich.edu}

\author{Nathan Kim}
\email{ncyk@umich.edu}
\orcid{0000-0002-9558-3530}

\author{Francesca Lameiro}
\email{flameiro@umich.edu}
\orcid{0009-0000-6961-7779}

\author{Wells Lucas Santo}
\email{ephemera@umich.edu}
\orcid{0009-0000-0748-0105}

\author{Shreya Chowdhary}
\email{schowdha@umich.edu}
\orcid{0009-0004-1799-7836}

\author{Eugene Kim}
\email{genakim@umich.edu}
\orcid{0009-0004-8056-1103}

\author{Justine Zhang}
\email{tisjune@umich.edu}
\orcid{0000-0003-1553-741X}

\affiliation{
  \institution{University of Michigan}
  \city{Ann Arbor}
  \state{Michigan}
  \country{USA}
}

\renewcommand{\shortauthors}{Zheng et al.}

\begin{abstract}
In the face of increasing austerity and threats of AI-enabled labor replacement at the University of Michigan, a group of workers and students have coalesced around the project of “AI resistance” since Fall 2024. Forming a cross-departmental coalition including librarians, faculty, staff, graduate workers, and undergraduate students, we have hosted a public workshop questioning the techno-deterministic inevitability of AI use at the University and are working with other campus organizations to maintain an ongoing organizing space. This workshop submission incorporates our reflections thus far on the strategies we’ve employed, the challenges to collective resistance, and our role as workers in resisting AI within the University. Our aim for this work is to provide concrete inspiration for technologists, students, and staff looking to resist AI techno-solutionism within their own universities.

\end{abstract}



\maketitle

\section{AI Technosolutionism at the University}

In September 2024, OpenAI CEO Sam Altman visited the University of Michigan to speak to hundreds of students, faculty, and staff about ChatGPT, which he claimed constituted “the most exciting piece of the technological revolution in human history” \cite{altman2024michigan}. While Altman’s visit was boasted about in several university publications, less publicized were the University’s commitments that year totaling \$198.7 million to three venture capital funds personally overseen by Altman himself. Altman was hosted by Daniel Feder, Senior Managing Director of the University of Michigan endowment, who argued to Fortune several months prior that the hundreds of millions sent to Altman’s funds that year were “an extension of ongoing investment strategies” \cite{mathews_exclusive_2023}. Feder referred specifically to the endowment’s investments, but the investment into AI through both public spectacle and the University’s endowment is consistent with the University’s strategy to invest broadly in AI as a solution for all aspects of life and labor within the university \cite{members_of_the_tahrir_coalition_university_2024}. 

In the same year, the University entered into a contract with Microsoft to develop a generative AI chatbot that is now publicly available on the Google Play Store and Apple App Store, promising “limitless potential” through the chatbot like showing the day’s lunch menu (which was previously freely accessible on several university websites) \cite{information_and_technology_services_go_2025}. The School of Information (SI) and the Department of Computer Science and Engineering (CSE) are now designing new degree programs on "applied AI," receive AI research grants from organizations like OpenAI and Google \cite{funder_tracer_2024}, and produce research papers that legitimize generative AI proliferation under the banner of "human-centered AI."

As a result of these decisions by University deans and administrators, many of our fellow workers have found their work to be in alignment with the AI hype machine and are not agitated towards resisting the project of AI. At the same time, it is exactly our position as workers that reveals what kinds of problems the University attempts to solve through its investments into AI. The organizing work that members of our group are involved with informs our analytical approach—we know that the University is developing an in-house “GraderGPT” and pushes faculty to break strikes to discipline workers \cite{graderGPT, um_admin_fabricate_grades_2023}. AI, for University administrators, is a solution to the “problems” of worker-led fights for living wages and sensible working conditions \cite{kornbluh_academes_2020, seybold_jason_2023, kirschenbaum_ai_2024}. 

\section{Resisting AI as Workers and Students}

In November 2024, over 50 University of Michigan workers — including faculty, staff, librarians, and graduate students — participated in a half-day workshop titled, “AI is Not Inevitable: A Workshop for AI Skeptics at U-M.” This event was organized by a subset of these workers, who are all authors on this workshop submission and are disturbed by the University’s development of custom generative AI applications (such as “U-M GPT”), some of which have been marketed as a replacement for teaching labor. Our mini ‘AI Workers Inquiry’ activity created a space for participants to compare their experiences as workers in an increasingly AI-forward University and how the University’s investments in AI are impacting them. The 9 questions in our inquiry—inspired by Marx’s 100 question-long “workers' inquiry,” a method for workers’ collective sense-making about their own working conditions and means of resistance \cite{marx_workers_1880, miceli_data_2024}—helped us interrogate the perceived value of our work, the impacts of generative AI, and the rationale/motivation for intensified AI investment \footnote{See our AI worker's inquiry here: \url{https://bit.ly/AIworkersinquiry4edu}}.

Following the initial workshop, participants have continued to meet monthly to organize against AI at the University. Through these meetings, we aim to create a space for University staff, faculty, students, and community members to meet with others expressing doubts about the unrelenting push for generative AI in higher education. Presently, the coalition is working on a variety of interventions, including: (a) collectively developing resources and talking points against the use of AI in higher education, (b) “art against AI” projects, and (c) collaborations with other forces on campus like campus unions who share a common goal of fighting against the corporatization (and now outright fascistization) of higher education \cite{graeber2014anthropology}.

\section{Challenges to Organizing an AI Resistance}

All of us (the authors of this paper) work on technology, either as researchers studying sociotechnical systems or as staff in the IT department at the university. This positionality places us in a core hub of the University’s techno-fascist project, which affords us a particularly direct view into the project but also introduces complications in our efforts to resist. As technologists critical of AI, we find ourselves acting counter-culturally against the culture of disengagement and AI hype that many of our coworkers have been conditioned into. Our immediate workplaces foster an ideology of depoliticization and meritocracy that frames the university’s techno-fascist project as intrinsically progressive \cite{newAestheticsOfFascism, mcquillan2022resisting, golumbia2009cultural}. In this section, we will discuss the concrete challenges to organizing an AI resistance.

First, we have struggled to understand the impacts of “AI” within our university. Though coalition members can provide many personal anecdotes about their encounters with 
the university's AI rollout,
a comprehensive and systematic understanding of its costs and outcomes is not immediately within our grasp. Part of the issue stems from obfuscation on the institution’s part \cite{ahmed2021complaint}, with University-produced materials narrowly focusing on AI as an enclosed technical system. Our questions—what the internal and externalized costs of these AI systems are, whether students and staff find these systems useful, and what University resources have been deployed for the development of AI tools—fall out of scope for administrators focused on developing new streams of revenue through external corporate partnerships. Our coalition suggests that future work will 
require more intensive oppositional research tactics, like FOIA requests.

Second, we have struggled to resist the dominant culture of our workplaces, experiencing opposition from our fellow coworkers. After posting flyers for our “AI is Not Inevitable” workshops across the School of Information, we were surprised to find satirical counter-flyers proclaiming "The ‘Internet’ is Not Inevitable,” placed next to our flyers the following morning.\footnote{See figure 1 in the appendix for our flyers and counter-flyers} Within Information and Technology Services (a central IT organization that manages enterprise services at U-M), there is a heavy-handed management-led propaganda push that generative AI is the future and resistance is futile, perpetuating long-standing paternalistic dynamics between ITS and the University community. Individual staff members have doubts though, and our coalition is organizing around these doubts. The new staff union University Staff United– containing library workers, academic advisors, tech workers, admin staff, and many others– has also served as a well of support for AI skepticism.

Finally, we have encountered difficulties around developing a shared understanding of what the term “AI” itself signifies. This lack of understanding about the impacts of “AI” at the University might be understood as linked to lack of specificity that is embedded within the term itself. One member of our coalition suggested that we should focus on critiquing “generative AI” technologies specifically, noting the many socially beneficial uses of AI, like cancer detection. Many members felt unconfident critiquing AI due to their non-technologist backgrounds. This lack of conceptual clarity about AI is what gives the term “AI” its political potency \cite{suchman2023uncontroversial, kirschenbaum_ai_2024}. Lucy Suchman describes the function of “AI” as a “floating signifier,” or “a term that suggests a specific referent, but works to escape definition in order to maximize its suggestive power” \cite{suchman2023uncontroversial}. In response, we are reminded of the necessity of analyzing the University’s AI projects in the context of the power relations between the University and the worker in a world structured by interconnecting axes of oppression \cite{collins2022black, tacheva2023ai}; in our organizing, we turn towards the concrete and material impacts of the university’s AI projects, rather than their technical or design intricacies \cite{aliDefiningAI}, to understand AI as one part of the University's broader neoliberal political agenda \cite{harney2013undercommons, eaton2022bankers, lin_techniques_2021,bousquet2017university}.

\section{Resisting AI is Resisting the Neoliberal University}

Resisting “AI” within the University feels like fighting an apparition or an idea. It feels that we need to organize on two different, parallel fronts. First, by becoming more concrete and specific in our resistance, focusing on specific products and instantiations of AI, such as automated grading technologies. Second, by becoming more abstract, engaging in our own language games, 
re-engineering the referents for “AI” through art and propaganda, in order to explicitly link the project of “AI” to techno-fascist impulses towards centralized control and austerity \cite{golumbia2013cyberlibertarianism, mcelroy2024silicon}. 

Our organizing efforts have brought us into contact with other realities within our nascent AI University: fellow "AI skeptics" have reported shortages in staffing, students are creating literature reviews filled with hallucinated sources, and physical facilities in humanities departments remain neglected while the University invests in AI technosolutions \cite{limitsOfRefusal, memosBloodFire}. 
We must advance an understanding of “AI,” not as an amorphous inevitability, but as a series of concrete decisions made by the University that advance a project of austerity and labor devaluation \cite{hong_strategic_2024, austerityIsClassWar}.

While the options legible to us as technologists or HCI scholars might be sharper critique or more conscientious design, the stakes for us as workers point toward a different set of practices for resistance: mapping the University’s financial investments, critically examining the University’s AI projects and investments, and documenting material impacts on workers, students, and the environment. Simultaneously, we must create space for workers and students to collectively articulate their desires for better universities, workplaces, and worlds—and contest the ways that “AI” shrinks these desires to a point.

\newpage

\bibliography{references.bib}

\newpage
\appendix
\section{Appendix}

\begin{figure}[h]
  \centering
  \includegraphics[width=0.4\hsize]{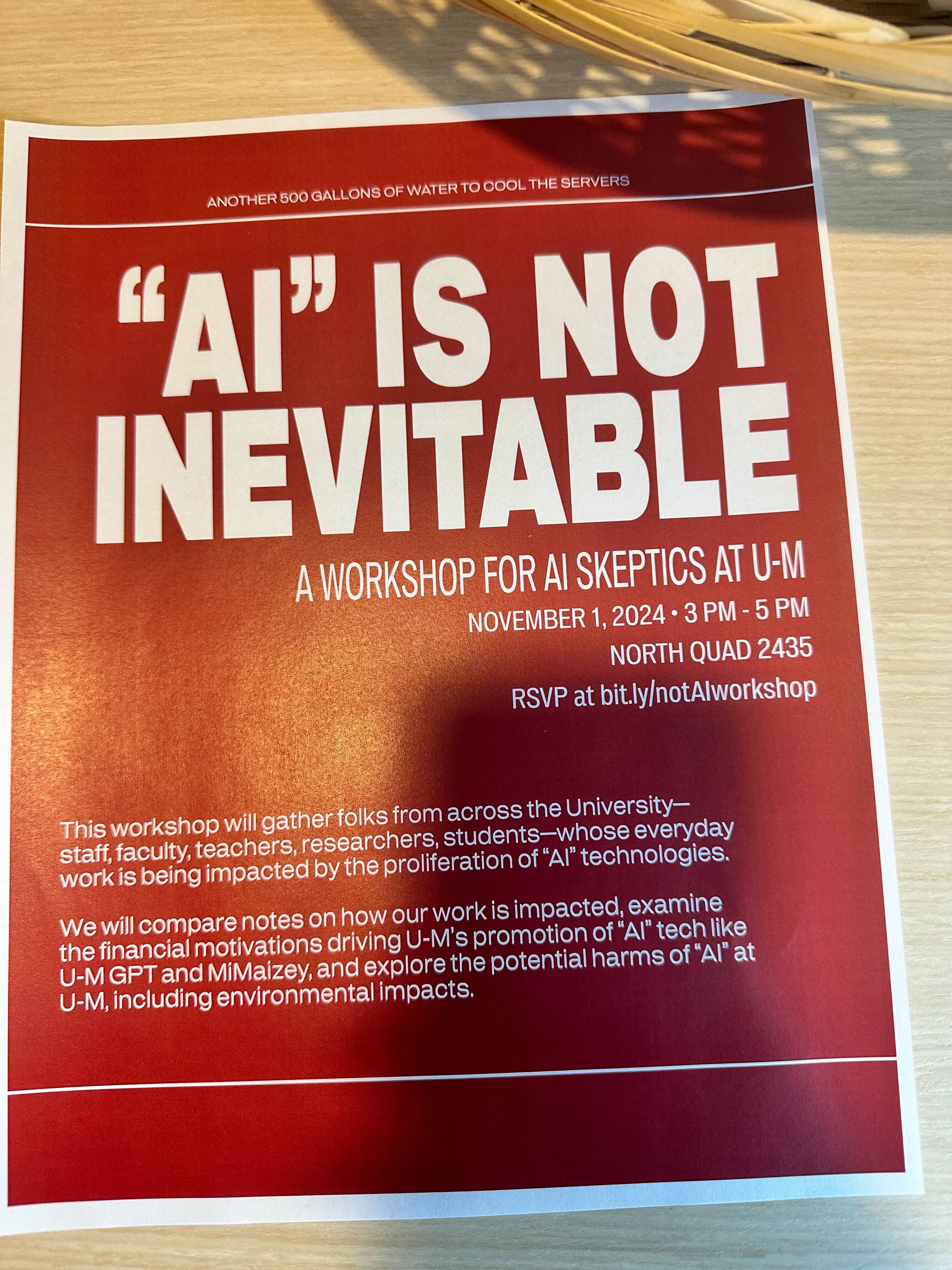}
  \includegraphics[width=0.4\hsize]{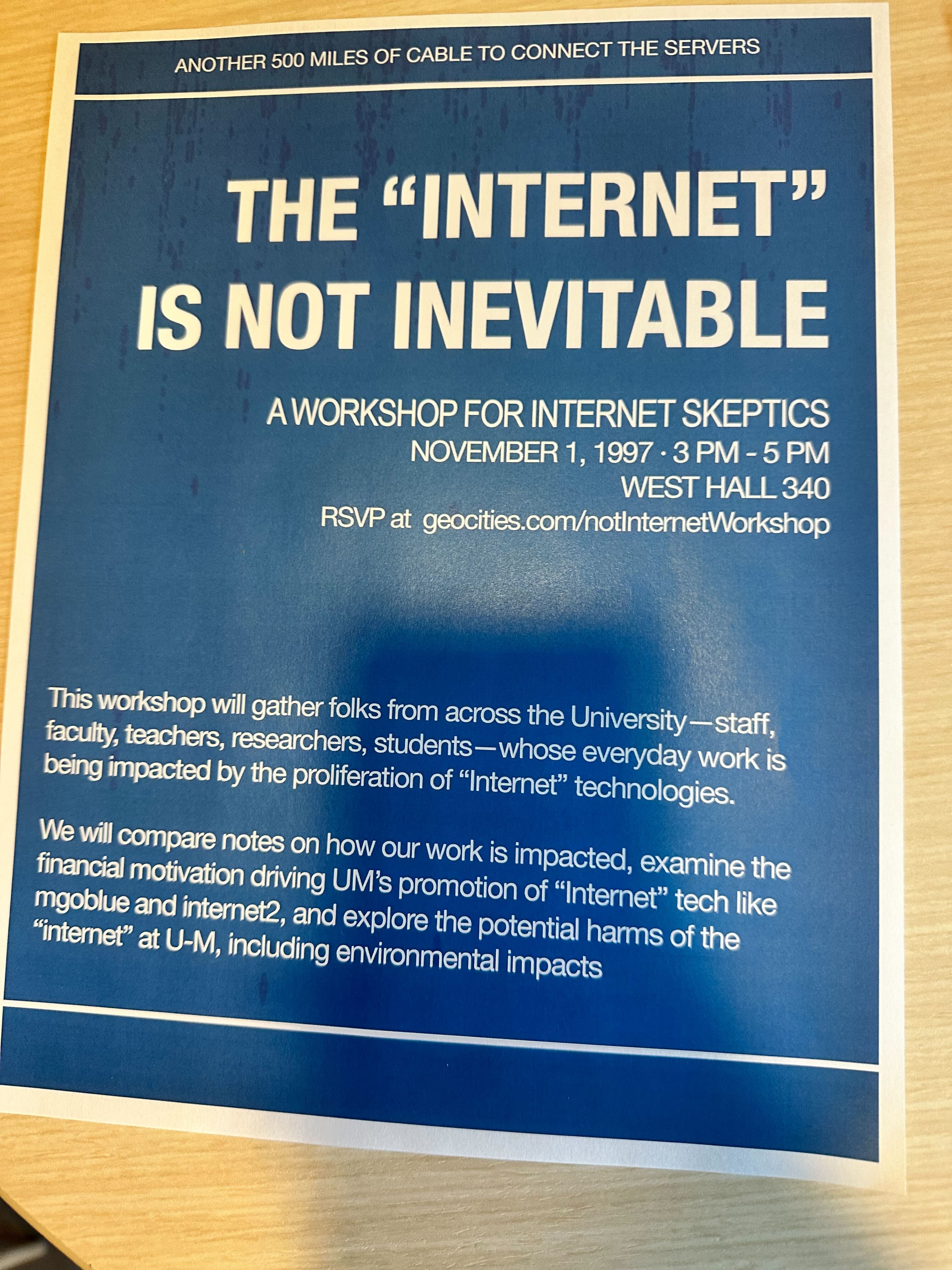}
  \caption{Left: our group's promotional flyer for our workers' inquiry workshop; Right: satirical counter-flyer}
  \Description{Two flyers at the School of Information.}
  \label{fig:flyers}
\end{figure}

\bibliographystyle{ACM-Reference-Format}
\end{document}